\documentclass{article}
\usepackage{graphics}
\usepackage{epsfig}

\begin{document}

\begin{center} {\large \bf Polarization properties of 'slow' light}
\end{center}
\bigskip
\begin{center}
\textbf{V. A. Reshetov, I. V. Meleshko}\\
\bigskip
\textit{Department of General and Theoretical Physics, Tolyatti
State University, 14 Belorousskaya Street, 446667 Tolyatti, Russia}
\end{center}

\begin{abstract}
The propagation of the arbitrarily polarized pulse of the weak probe
field through the resonant medium of $\Lambda$-type three-level
atoms with degenerate levels adiabatically driven by the coherent
coupling field is considered. It is shown that such pulse is
decomposed in the medium into two orthogonally polarized dark-state
polaritons propagating with different group velocities. The
expressions for the polarizations and group velocities of these two
polaritons are obtained. The dependence of these polarizations and
group velocities on the values of the angular momenta of resonant
levels, on the polarization of the coupling field and on the initial
atomic state is studied.
\end{abstract}

\section{Introduction}

The remarkable reduction of the group velocity of light pulses
\cite{z1,z2} based on the phenomenon of the electromagnetically
induced transparency (EIT) \cite{z3,z4} provided a number of
applications, the most promising among them being the implementation
of quantum memory \cite{z5,z6,z7}. The recent experiments
\cite{z8,z9,z10} on EIT-based quantum memory demonstrate the
continuously enhancing memory efficiency and fidelity, bringing it
close to practical applications. Such memory is based on the concept
of dark-state polaritons in the three-level $\Lambda$-type systems,
proposed in \cite{z11,z12} and soon realized in rubidium vapor in
the experiment \cite{z13}. The group velocity of such polaritons may
be controlled by the adiabatically varying intensity of the driving
field to store single-photon pulses in the resonant media or to
retrieve them. The most natural way for q-bit encoding is provided
by the photon two polarization degrees of freedom, as it was
implemented in the experiments \cite{z9,z10}. However the group
velocity of the dark-state polariton may depend essentially on its
polarization due to the optical anisotropy induced by the
polarization of the driving field, while for the effective storage
of the photon polarization q-bit its both polarization components
must be stopped in the medium simultaneously. The objective of the
present paper is to study the polarization properties of the
dark-state polaritons formed in the three-level $\Lambda$-type
systems with degenerate levels, which are in many experiments the
hyperfine structure components of alkali atoms degenerate in the
projections of the atomic total angular momentum on the quantization
axis.

\section{Basic equations and relations}

We consider the pulse of the weak probe field propagating along the
sample axis $Z$ with the carrier frequency $\omega$, which is in
resonance with the frequency $\omega_{0}$ of an optically allowed
transition $J_{a}\rightarrow J_{c}$ between the ground state $J_{a}$
and the excited state $J_{c}$, while the strong coherent coupling
field propagates in the same direction with the carrier frequency
$\omega_{c}$, which is in resonance with the frequency $\omega_{c0}$
of an optically allowed transition $J_{b}\rightarrow J_{c}$ between
the long-lived state $J_{b}$ and the same excited state $J_{c}$
(Fig.1). Here $J_{a}$, $J_{b}$ and $J_{c}$ are the values of the
angular momenta of the levels. The electric field strength of the
coupling field and that of the probe field may be put down as
follows:
     \begin{equation}\label{q1}
\textbf{E}_{c}=e_{c}(t-z/c)\textbf{l}_{c} e^{-i\omega_{c}(t-z/c)}+
c.c.,
     \end{equation}
     \begin{equation}\label{q2}
\textbf{E}=\textbf{e}(t,z) e^{-i\omega (t-z/c)}+c.c.,
     \end{equation}
where $e_{c}$ is the slowly-varying amplitude of the coupling field
and $\textbf{l}_{c}$ is its constant unit polarization vector, while
$\textbf{e}$ is the slowly-varying vector amplitude of the probe
pulse, which satisfies the equation:
    \begin{equation}\label{q3}
 \left(\frac{\partial}{\partial t} + c \frac{\partial}{\partial
 z}\right) \textbf{e} = \frac{i \omega n_{0}|d|}{2\varepsilon_{0}}
 tr\left\{\hat{\textbf{g}}\hat{\rho}\right\},
     \end{equation}
as it follows from Maxwell equations, while the evolution of the
atomic slowly-varying density matrix $\hat{\rho}$ in the
rotating-wave approximation is described by the equation:
     \begin{equation}\label{q4}
 \frac{\partial \hat{\rho}}{\partial t} =
 \frac{i}{2}\left[\hat{V},\hat{\rho}\right] +
 \left(\frac{d\hat{\rho}}{dt}
 \right)_{rel},
     \end{equation}
     \begin{equation}\label{q5}
\hat{V} = 2(\Delta\hat{P}_{c}+\delta\hat{P}_{b}) + \hat{D} +
\hat{D}^{\dag},~\hat{D} = \Omega_{c} \hat{g}_{c} + \hat{G},
     \end{equation}
     \begin{equation}\label{q6}
\hat{g}_{c}=\hat{\textbf{g}}_{c}\textbf{l}_{c}^{*},~ \hat{G} =
(2|d|/\hbar)\hat{\textbf{g}}\textbf{e}^{*}.
     \end{equation}
     \begin{figure}[t]\center
\includegraphics[width=7cm]{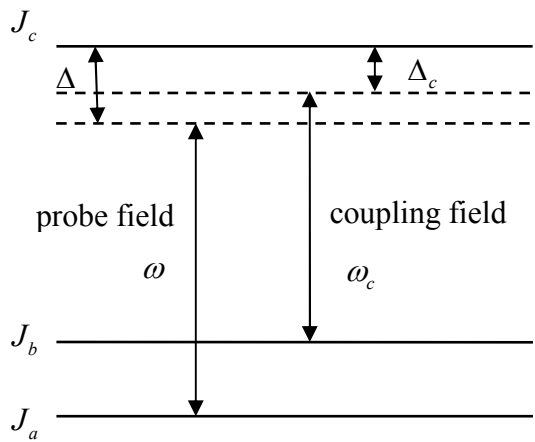}
\caption{The level diagram.}
     \end{figure}
Here $n_{0}$ is the concentration of resonant atoms,
$\hat{\textbf{g}}$ and $\hat{\textbf{g}}_{c}$ are the dimensionless
electric dipole moment operators for the transitions
$J_{a}\rightarrow J_{c}$ and $J_{b}\rightarrow J_{c}$,
$d=d(J_{a}J_{c})$ and $d_{c}=d(J_{b}J_{c})$ being the reduced matrix
elements of the electric dipole moment operators for these
transitions, $\Delta=\omega-\omega_{0}$ and
$\Delta_{c}=\omega_{c}-\omega_{c0}$ are the frequency detunings from
resonance of the probe and of the coupling fields, while
$\delta=\Delta-\Delta_{c}$, $\hat{P}_{\alpha}$ is the projector on
the subspace of the atomic level $J_{\alpha}$ ($\alpha=a,b,c$),
$\Omega_{c}=2|d_{c}|e_{c}/\hbar$ is the reduced Rabi frequency for
the coupling field. The matrix elements of the circular components
$\hat{g}_{q}$ and $\hat{g}_{cq}$ ($q=0,\pm 1$) of vector operators
$\hat{\textbf{g}}$ and $\hat{\textbf{g}}_{c}$ are expressed through
Wigner 3J-symbols \cite{z14}:
    \begin{equation}\label{q7}
  (\hat{g}_{q})^{ac}_{m_{a}m_{c}}=
(-1)^{J_{a}-m_{a}}\left(\matrix{J_{a}&1&J_{c}  \cr
-m_{a}&q&m_{c}}\right),
    \end{equation}
    \begin{equation}\label{q8}
  (\hat{g}_{cq})^{bc}_{m_{b}m_{c}}=
(-1)^{J_{b}-m_{b}}\left(\matrix{J_{b}&1&J_{c}  \cr
-m_{b}&q&m_{c}}\right).
    \end{equation}
Finally, the term $(d\hat{\rho}/dt)_{rel}$ in the equation
(\ref{q3}) describes the irreversible relaxation. Initially the
atoms are at the ground state $a$ the atomic density matrix being
$$\hat{\rho}(0)=\hat{\rho}_{a}.$$
In the linear approximation for the probe field we obtain from the
equations (\ref{q4})-(\ref{q5}) for the elements of the atomic
density matrix the following equations:
    \begin{equation}\label{q9}
\left(\frac{\partial}{\partial t} + \gamma-i\Delta\right)
\hat{\rho}^{ca} = \frac{i}{2}
\left(\Omega_{c}\hat{g}_{c}^{\dag}\hat{\rho}^{ba} +
\hat{G}^{\dag}\hat{\rho}_{a}\right),
    \end{equation}
    \begin{equation}\label{q10}
\left(\frac{\partial}{\partial t} + \Gamma-i\delta \right)
\hat{\rho}^{ba} = \frac{i}{2} \Omega_{c}\hat{g}_{c}\hat{\rho}^{ca},
    \end{equation}
where
$$\hat{\rho}^{\alpha\beta}= \hat{P}_{\alpha}\hat{\rho}
\hat{P}_{\beta},~ \alpha,\beta=a,b,c,$$ while the irreversible
relaxation is simply characterized by the two real relaxation rates
-- $\gamma$ for the optically allowed transition $J_{a}\rightarrow
J_{c}$ and $\Gamma$ for the optically forbidden transition
$J_{a}\rightarrow J_{b}$:
$$ \left(\frac{d\hat{\rho}}{dt} \right)_{rel}^{ca} = - \gamma
\hat{\rho}^{ca},~  \left(\frac{d\hat{\rho}}{dt} \right)_{rel}^{ba} =
- \Gamma \hat{\rho}^{ba}.$$ In the adiabatic approximation, when the
coupling field varies slowly:
$$\gamma T \gg 1,~~~\Omega_{c}^{2}T \gg \gamma,$$
$T$ being the characteristic time scale, while the relaxation at the
forbidden transition remains negligible $$\Gamma T \ll 1,$$ in the
case of single-photon and two-photon resonances $$\Delta \preceq
\gamma,~~~\delta T \ll 1,$$  from (\ref{q9})-(\ref{q10}) it follows:
    \begin{equation}\label{q11}
\Omega_{c}\hat{g}_{c}^{\dag}\hat{\rho}^{ba} +
\hat{G}^{\dag}\hat{\rho}_{a}=0,
    \end{equation}
    \begin{equation}\label{q12}
\frac{\partial \hat{\rho}^{ba}}{\partial t}  = \frac{i}{2}
\Omega_{c}\hat{g}_{c}\hat{\rho}^{ca}.
    \end{equation}
The equation (\ref{q11}), which is the approximation of the relation
$$\hat{D}^{\dag}\hat{\rho}=0,$$ linear in the probe field,
means that only the dark states contribute to the solution of the
equation (\ref{q12}). In order to express $\hat{\rho}^{ca}$ through
$\hat{\rho}^{ba}$ we multiply both parts of the equation (\ref{q12})
by the matrix $\hat{g}_{c}^{\dag}$ from the left and consider the
orthonormal set of eigenvectors $|c_{n}>$ of the hermitian operator
$\hat{g}_{c}^{\dag}\hat{g}_{c}$ acting at the subspace of the
excited level $c$:
    \begin{equation}\label{q13}
\hat{g}_{c}^{\dag}\hat{g}_{c}|c_{n}> = c_{n}^{2}|c_{n}>,
n=1,...,2J_{c}+1,
    \end{equation}
the corresponding eigenvalues being non-negative $c_{n}^{2}\geq 0.$
Then, after multiplying both parts of the equation (\ref{q12}) from
the left by the matrix
    \begin{equation}\label{q14}
\hat{D}_{c}=\sum_{n}\frac{1}{c_{n}^{2}}|c_{n}><c_{n}|,
    \end{equation}
we obtain
    \begin{equation}\label{q15}
\hat{P}_{c}^{b}\hat{\rho}^{ca} = -\frac{2i}{\Omega_{c}}
\frac{\partial }{\partial t}\left(\hat{D}_{c}
\hat{g}_{c}^{\dag}\hat{\rho}^{ba}\right),
    \end{equation}
where
    \begin{equation}\label{q16}
\hat{P}_{c}^{b}=\sum_{n}|c_{n}><c_{n}|,
    \end{equation}
while the summation in the equations (\ref{q14}) and (\ref{q16}) is
carried out only over eigenvectors $|c_{n}>$ with non-zero
eigenvalues $c_{n}^{2}>0$, $\hat{P}_{c}^{b}$ being the projector on
the subspace formed by such eigenvectors. The eigenvectors $|c_{n}>$
with zero eigenvalues $c_{n}^{2}=0$ are not affected by the coupling
field and may be neglected under the assumed approximation. Now let
us introduce the vector field
    \begin{equation}\label{q17}
\textbf{p} = tr\left\{\hat{\textbf{g}}
\hat{D}_{c}\hat{g}_{c}^{\dag}\hat{\rho}^{ba}\right\},
    \end{equation}
describing the induced coherence at the forbidden transition
$J_{a}\rightarrow J_{b}$. With the two orthonormal vectors
$\textbf{l}_{i}$ in the polarization plane $XY$
($\textbf{l}_{j}\textbf{l}_{k}^{*}=\delta_{jk},~j,k=1,2$) we obtain
from (\ref{q3}), (\ref{q15}) and (\ref{q11}) for the components
$e_{k}=\textbf{e}\textbf{l}_{k}^{*}$ and
$p_{k}=\textbf{p}\textbf{l}_{k}^{*}$ the following equations:
     \begin{equation}\label{q18}
 \left(\frac{\partial}{\partial t} + c \frac{\partial}{\partial
 z}\right) e_{k} = \frac{\omega n_{0}|d|}{\varepsilon_{0}\Omega_{c}}
 \frac{\partial p_{k}}{\partial t},
     \end{equation}
    \begin{equation}\label{q19}
p_{k}=-\frac{2|d|}{\hbar\Omega_{c}}\sum_{j}a_{kj}e_{j},
    \end{equation}
    \begin{equation}\label{q20}
a_{kj}=
tr\left\{\hat{\rho}_{a}\hat{g}_{k}\hat{D}_{c}\hat{g}_{j}^{\dag}
\right\},~ \hat{g}_{k}=\hat{\textbf{g}}\textbf{l}_{k}^{*}.
    \end{equation}
Now let us choose the two orthonormal vectors $\textbf{l}_{i}$ as
the two eigenvectors of the hermitian $2\times 2$ matrix $a_{jk}$,
defined by (\ref{q20}). Then
    \begin{equation}\label{q21}
a_{kj}= a_{k}\delta_{k_{j}},
    \end{equation}
where $a_{k}$ ($k=1,2$) are the two real eigenvalues of this matrix.
By introducing via canonical transformation the field of the
dark-state polariton  \cite{z12}:
    \begin{equation}\label{q22}
\Psi_{k}= \cos \theta_{k} e_{k} - \sin \theta_{k} \lambda_{k} p_{k},
    \end{equation}
where
    \begin{equation}\label{q23}
\lambda_{k} = \sqrt{\frac{\hbar \omega
n_{0}}{2\varepsilon_{0}a_{k}}},
    \end{equation}
and the angle $\theta_{k}$ is determined by the equation
    \begin{equation}\label{q24}
\tan \theta_{k} = \frac{2|d|a_{k}\lambda_{k}}{\hbar \Omega_{c}},
    \end{equation}
neglecting the retardation of the coupling field
$\Omega_{c}(t-z/c)\simeq \Omega_{c}(t)$, we obtain from
(\ref{q18})-(\ref{q24}) the following equation:
     \begin{equation}\label{q25}
 \frac{\partial \Psi_{k}}{\partial t} + c \cos^{2} \theta_{k}
 \frac{\partial \Psi_{k}} {\partial z} = 0,
     \end{equation}
which describes the propagation of the dark-state polariton with the
group velocity
     \begin{equation}\label{q26}
V^{gr}_{k}=c \cos^{2}\theta_{k}=\frac{c}{1+n^{gr}_{k}},
     \end{equation}
     \begin{equation}\label{q27}
n^{gr}_{k}= \tan^{2}\theta_{k}=\frac{2|d|^{2}n_{0}\omega
a_{k}}{\hbar\Omega_{c}^{2}\varepsilon_{0}}.
     \end{equation}

\section{Discussion}

As it follows from (\ref{q22})-(\ref{q27}), the arbitrarily
polarized pulse of the weak probe field is decomposed under the
action of the strong coupling field into two components polarized
along the two eigenvectors of tensor (\ref{q20}) propagating with
different group velocities, which difference is determined by the
difference in the two eigenvalues of tensor (\ref{q20}). So the
polarization properties of the dark-state polaritons are totally
determined by the hermitian $2\times 2$ tensor $a_{jk}$ defined by
the equation (\ref{q20}). This tensor in its turn is determined by
the values of the angular momenta of resonant levels, by the
polarization of the coupling field and by the initial atomic state.
Let us now calculate the eigenvalues $a_{1}$ and $a_{2}$ and the
corresponding eigenvectors $\textbf{l}_{1}$ and $\textbf{l}_{2}$ of
tensor $a_{jk}$ for some transitions $J_{a}\rightarrow
J_{c}\rightarrow J_{b}$ with the angular momenta characteristic for
the experiments on the hyperfine structure components of atomic
levels. In the case of equilibrium initial atomic state and linearly
polarized coupling field
$$\hat{\rho}_{a}=\frac{\hat{P}_{a}}{2J_{a}+1},
~\textbf{l}_{c}=\textbf{l}_{x},$$ we obtain:
$$a_{1}=2,~a_{2}=0,~\textbf{l}_{1}=\textbf{l}_{y},~
\textbf{l}_{2}=\textbf{l}_{x},$$ for transitions $J_{a}=0\rightarrow
J_{c}=1\rightarrow J_{b}=1$,
$$a_{1}=1.111,~a_{2}=0.972,~\textbf{l}_{1}=\textbf{l}_{x},~
\textbf{l}_{2}=\textbf{l}_{y},$$
for transitions $J_{a}=1\rightarrow J_{c}=1\rightarrow J_{b}=2$,
$$a_{1}=2,~a_{2}=1.5,~\textbf{l}_{1}=\textbf{l}_{x},~
\textbf{l}_{2}=\textbf{l}_{y},$$ for transitions $J_{a}=1\rightarrow
J_{c}=2\rightarrow J_{b}=2$,
$$a_{1}=1.295,~a_{2}=0.907,~\textbf{l}_{1}=\textbf{l}_{x},~
\textbf{l}_{2}=\textbf{l}_{y},$$ for transitions $J_{a}=2\rightarrow
J_{c}=2\rightarrow J_{b}=3$,
$$a_{1}=2.96,~a_{2}=1.787,~\textbf{l}_{1}=\textbf{l}_{x},~
\textbf{l}_{2}=\textbf{l}_{y},$$ for transitions $J_{a}=2\rightarrow
J_{c}=3\rightarrow J_{b}=3$,
$$a_{1}=1.445,~a_{2}=0.944,~\textbf{l}_{1}=\textbf{l}_{x},~
\textbf{l}_{2}=\textbf{l}_{y},$$ for transitions $J_{a}=3\rightarrow
J_{c}=3\rightarrow J_{b}=4$,
$$a_{1}=3.832,~a_{2}=2.151,~\textbf{l}_{1}=\textbf{l}_{x},~
\textbf{l}_{2}=\textbf{l}_{y},$$ for transitions $J_{a}=3\rightarrow
J_{c}=4\rightarrow J_{b}=4$.

In the case of equilibrium initial atomic state and circularly
polarized coupling field
$$\hat{\rho}_{a}=\frac{\hat{P}_{a}}{2J_{a}+1},
~\textbf{l}_{c}=\textbf{l}_{+1},$$ we obtain:
$$a_{1}=2,~a_{2}=0,~\textbf{l}_{1}=\textbf{l}_{+1},~
\textbf{l}_{2}=\textbf{l}_{-1},$$ for transitions
$J_{a}=0\rightarrow J_{c}=1\rightarrow J_{b}=1$,
$$a_{1}=2.222,~a_{2}=0.833,~\textbf{l}_{1}=\textbf{l}_{+1},~
\textbf{l}_{2}=\textbf{l}_{-1},$$ for transitions
$J_{a}=1\rightarrow J_{c}=1\rightarrow J_{b}=2$,
$$a_{1}=1.444,~a_{2}=0.611,~\textbf{l}_{1}=\textbf{l}_{+1},~
\textbf{l}_{2}=\textbf{l}_{-1},$$ for transitions
$J_{a}=1\rightarrow J_{c}=2\rightarrow J_{b}=2$,
$$a_{1}=2.59,~a_{2}=1.12,~\textbf{l}_{1}=\textbf{l}_{+1},~
\textbf{l}_{2}=\textbf{l}_{-1},$$ for transitions
$J_{a}=2\rightarrow J_{c}=2\rightarrow J_{b}=3$,
$$a_{1}=1.392,~a_{2}=0.832,~\textbf{l}_{1}=\textbf{l}_{+1},~
\textbf{l}_{2}=\textbf{l}_{-1},$$ for transitions
$J_{a}=2\rightarrow J_{c}=3\rightarrow J_{b}=3$,
$$a_{1}=2.89,~a_{2}=1.362,~\textbf{l}_{1}=\textbf{l}_{+},~
\textbf{l}_{2}=\textbf{l}_{-1},$$ for transitions
$J_{a}=3\rightarrow J_{c}=3\rightarrow J_{b}=4$,
$$a_{1}=1.402,~a_{2}=0.971,~\textbf{l}_{1}=\textbf{l}_{+1},~
\textbf{l}_{2}=\textbf{l}_{-1},$$ for transitions
$J_{a}=3\rightarrow J_{c}=4\rightarrow J_{b}=4$.

Here $\textbf{l}_{x}$ and $\textbf{l}_{y}$ denote the unit vectors
of the Cartesian axes, while
$$\textbf{l}_{\pm 1}= \mp\frac{\textbf{l}_{x}\pm
i\textbf{l}_{y}}{\sqrt{2}}$$ are the two circular vectors.

In the experiment \cite{z9} the coupling field was linearly
polarized in the direction of propagation of the probe field:
$\textbf{l}_{c}=\textbf{l}_{z}$ ($\pi$-polarized), while it
propagated in the perpendicular direction, and the atoms were
prepared at the pure Zeeman state with the zero projection on the
quantization axis. Under such conditions the group velocity of the
probe pulse does not depend on its polarization for the arbitrary
value of angular momenta, as it follows from (\ref{q20}). The same
remains true for the unprepared atoms, which are initially in the
equilibrium state with equally populated Zeeman sublevels, as it
also follows from (\ref{q20}).

\section{Conclusions}

In the present article we consider the propagation of the
arbitrarily polarized pulse of the weak probe field through the
resonant medium of $\Lambda$-type three-level atoms with degenerate
levels adiabatically driven by the coherent coupling field. It is
shown that such pulse is decomposed in the medium into two
orthogonally polarized dark-state polaritons propagating with
different group velocities. The polarizations of these two
polaritons and the difference in their group velocities are
determined by the values of the angular momenta of resonant levels,
by the polarization of the coupling field and by the initial atomic
state.

{\bf Acknowledgements}

Authors are indebted for financial support of this work to Russian
Ministry of Science and Education (grant 2.2407.2011).

\end{document}